\documentclass[twocolumn]{aastex63}
\usepackage{hyperref}
\usepackage{amsmath,amssymb,mathrsfs}
\usepackage{soul}
\usepackage{lineno}
\usepackage{psfig}
\usepackage{graphicx}

\def\ibh {OGLE-2011-BLG-0462}

\def \xmm {\emph{XMM-Newton}}

\def \cha {\emph{Chandra}}

\def \mdot {\dot M}

\def\msun{{\rm M}_{\odot}}

\def\lsun{L_{\odot}}

\def\ltsima{$\; \buildrel < \over \sim \;$}
\def\lsim{\lower.5ex\hbox{\ltsima}}
\def\gtsima{$\; \buildrel > \over \sim \;$}
\def\gsim{\lower.5ex\hbox{\gtsima}}

\graphicspath{{fig/}}

\defcitealias{Davies81}{DP81} 

\received{April 21, 2022 }
\revised{ June 13, 2022}
\accepted{June 14, 2022 }

\shorttitle{X-ray observations of collapsed objects candidates found with microlensing  }

\begin{document}

\title{X-ray observations of the isolated black hole candidate \ibh , and other collapsed objects discovered through gravitational microlensing}

\correspondingauthor{ }
\email{sandro.mereghetti@inaf.it}

\author[0000-0003-3259-7801]{S.~Mereghetti}
\affiliation{INAF, Istituto di Astrofisica Spaziale e Fisica Cosmica, via A. Corti 12, I-20133 Milano, Italy}
 
\author[0000-0000-0000-0000]{L.~Sidoli}
\affiliation{INAF, Istituto di Astrofisica Spaziale e Fisica Cosmica, via A. Corti 12, I-20133 Milano, Italy}

\author[0000-0000-0000-0000]{G.~Ponti}
\affiliation{INAF, Osservatorio Astronomico di Brera, via E. Bianchi 46, I-23807 Merate (LC), Italy  }
  \affiliation{Max-Planck-Institut f{\"u}r extraterrestrische Physik, Giessenbachstrasse, D-85748, Garching, Germany}
  
\author[0000-0002-0653-6207]{A.~Treves}
\affiliation{Universit\`a dell’Insubria, via Valleggio 11, I-22100 Como, Italy }
\affiliation{INAF, Osservatorio Astronomico di Brera, via E. Bianchi 46, I-23807 Merate (LC), Italy  }

  
\begin{abstract}
Isolated black holes and neutron stars can be revealed through the observation of long duration gravitational microlensing events.  A few candidates have been found in surveys of stars in the direction of the Galactic bulge. Recently,  thanks to the addition of astrometric information at milliarcsecond level, it has been possible to reduce the uncertainties in the masses and distances for some of these ``dark'' gravitational lenses  and select the most promising candidates. 
These isolated compact objects might emit X-rays powered by accretion from the interstellar medium.  
Using data of the   \cha ,  \xmm , and INTEGRAL  satellites, we searched for X-ray emission in 
 the isolated black hole candidate \ibh, and in several other putative collapsed objects found with gravitational microlensing.   
  \ibh\  has been recently interpreted as a 7.1 $\msun$ black hole at a distance of 1.6  kpc, although a different group obtained a mass range (1.6-4.4 $\msun$) that cannot exclude a massive neutron star.  
We have derived  upper limits on the flux from \ibh\ of  9$\times10^{-15}$ erg cm$^{-2}$ s$^{-1}$ in the  0.5-7 keV range and $\sim2\times10^{-12}$ erg cm$^{-2}$ s$^{-1}$ in the 17-60 keV range.  The implied X-ray luminosity is consistent with the small radiative efficiency expected for a black hole and disfavour a neutron star interpretation.  Limits down to a factor about five lower are obtained for the soft X-ray flux of other candidates, but their interpretation is affected by larger uncertainties in the masses, distances and spatial velocities.  
\end{abstract}

\keywords{X-ray sources,  Stellar mass black holes, Gravitational lensing, Bondi accretion}


\section{Introduction}
\label{sec:intro}

Our galaxy is thought to contain of the order of 10$^7$-10$^9$ black holes (BH), but only a few tens   have  been discovered so far  in X-ray binaries  (e.g., \citet{2016A&A...587A..61C}). In fact, X-ray emission powered by accretion of matter from the binary companion star allows us to select candidates, which can be eventually confirmed as BH by subsequent dynamical mass measurements. The presence of BH has been claimed also in a few non-interacting binaries, based on radial velocity measurements of their companions \citep{2019Natur.575..618L,2019Sci...366..637T,2021ApJ...913...48G}, but their true BH nature is debated (see, e.g., \citet{2020Sci...368.3282V,2021MNRAS.502.3436E}, and references therein).
A significant number of the BH in the Galaxy should be isolated,  either because they originated from single  stars or  because the binary in which they formed was disrupted by the natal supernova kick. 
In principle, such isolated BH can  be detected if they accrete matter from the interstellar medium (e.g., \citet{1971SvA....15..377S,1982ApJ...255..654I,1993A&A...277..477C,2002MNRAS.334..553A}), as it was also proposed for isolated neutron stars \citep{1970ApL.....6..179O}. 
However, several searches for isolated neutron stars (NS) and BH fed by accretion from the interstellar medium (ISM) gave negative results \citep{1995AJ....109.1199S,1999A&A...341L..51S,2003ApJ...596..437C,2006ApJS..165..173M}, despite the initially optimistic estimates on the number of detectable objects \citep{1991A&A...241..107T,1993ApJ...403..690B}. 

Isolated BH can also be detected through gravitational microlensing, when they pass in front of background stars \citep[e.g.,][]{1986ApJ...304....1P}.
Several long-duration microlensing events found in  surveys targeted to stars in the Galactic bulge should be due to BH, because the relative rarity of these objects is compensated by their large cross section.
For most microlensing events only the crossing time of the Einstein radius can be measured. This timescale depends not only on the lens mass, but also on the relative velocities and distances of the lens and the lensed star. Therefore, 
on the basis of the sole crossing time,  it is not possible to determine the individual values of these parameters.  
More information on the mass of the lens can be obtained in sufficiently long events, in which the effects caused by the non-linear motion of the Earth can be detected in the light curve of the magnified star \citep{1992ApJ...392..442G}. The first BH candidates selected with this method were reported by \citet{2002ApJ...579..639B}  and \citet{2002MNRAS.329..349M}. Note that, despite the  ``parallax microlensing" name of this phenomenon, it involves only photometric observations to measure the time-dependent deviations of the source magnification from the light curve expected with a simple model based on linear motions of observer, lens and lensed star. Due to the intrinsic degeneracy between the parameters that characterize parallax microlensing (in particular the one between lens mass and relative velocities) the masses derived in this way are subject to large uncertainties. This problem has usually been tackled by statistical analysis based on our best knowledge of the distributions of masses, positions and velocities of the lens and of the lensed star \citep{2002ApJ...576L.131A}. The nature of the lens (i.e. BH, NS,  white dwarf, normal star)  can thus be assessed in probabilistic terms. 

A step forward to reduce such a degeneracy can be done if also the deviations in the apparent position of the magnified star caused by the gravitational lens are detected. This requires very accurate astrometric measurements at the milliarcsecond level. 
Recently, using this ``astrometric microlensing'' method, \citet{2022arXiv220113296S} derived a   mass of  7.1$\pm$1.3 $\msun$  and  distance of 1.58$\pm$0.18 kpc for the lens object in \ibh . Based on the lack of detectable light from the lens, these authors concluded that this is the first unambiguous detection of an isolated stellar-mass BH.  Somewhat different results have been reported for the same event by \citet{2022arXiv220201903L}, who  inferred a lens mass in the range 1.6 - 4.4 $\msun$ which leaves open the possibility that \ibh\ be an isolated NS.  

X-ray observations of the compact objects found with gravitational microlensing can provide   useful constraints on the physics of accretion. This is possibly the only way to derive information on accretion of ISM onto isolated compact objects of stellar mass. 
A systematic comparative study between microlensing events and ROSAT sources was presented by  \citet{2012A&A...539A..52S}, finding no meaningful matching.
Up to now, deep searches for X-ray emission have been reported only for MACHO-96-BLG-5  \citep{2005ApJ...631L..65M,2006ApJ...651.1092N}. The negative results imply accretion on the putative BH at a rate lower than $\sim10^{-9}$ of the Eddington value. 

Here we report on  new and archival \cha\ observations of the  isolated BH candidate \ibh\ \footnote{ Our results supersede the limits quoted in \citet{2022arXiv220201903L}, which were obtained from a simple inspection of X-ray catalogues.}. We also carried out a search for X-ray emission using all the available \cha\ and \xmm\ archival data for several other  gravitational microlensing sources proposed,  with different levels of confidence, as  BH and NS candidates. 
To our knowledge, only the results for MACHO-98-BLG-5 have been previously reported in the literature
All the considered targets are briefly described in Section~\ref{sec-targets} while the data analysis and results are given in Section~\ref{sec-results}. The implications for the physics of accretion and future prospects are discussed in Sections~\ref{sec-disc} and ~\ref{sec-conc}.

\section{Candidate BH and NS found with gravitational  microlensing}
 \label{sec-targets}

\subsection{ \ibh\ }

\ibh\ (also known as MOA-2011-BLG-191) is probably the most extensively studied BH candidate found through astrometric microlensing.
Combining numerous data sets, including accurate astrometry carried out with the Hubble Space Telescope,  \citet{2022arXiv220113296S} derived a lens mass $M_L=7.1\pm1.3$ $\msun$.
They also obtained stringent limits on the lens luminosity that rule out a non-degenerate star. These authors could exclude the possibility that the lens is a close\footnote{ A wide binary could not be excluded.} binary system,   thus concluding that it is an isolated BH. The derived value of the lens distance is $D_L =1.58\pm0.18$ kpc,  only slightly dependent on the distance of the lensed star, and the lens transverse velocity is $V_T\sim$45 km s$^{-1}$. 
They also noted that the region surrounding the lens is affected by severe differential extinction, which may indicate
that locally the density of the ISM is large.

This microlensing event was studied also  by \citet{2022arXiv220201903L} who confirmed the dark nature of the lens. However,  depending on the relative weights given to the astrometric and photometric data used in the fits, these authors
found two  solutions leading to   different values for the lens mass, distance and velocity  (see Table~\ref{tab-sources}). {  While one   solution gives  $M_L=3.79^{+0.62}_{-0.57}$ $\msun$, hence a high probability for a BH, the other one gives $M_L=2.15^{+0.67}_{-0.54}$ $\msun$,  corresponding to   comparable NS and BH probabilities. 
In both solutions, the  transverse velocity of the lens ($V_T<$25 km s$^{-1}$) is smaller than that obtained by \citet{2022arXiv220113296S}.

\subsection{ MACHO-99-BLG-22 }

MACHO-99-BLG-22 (also known as OGLE-1999-BUL-32) was first pointed out as a possible BH by \citet{2002MNRAS.329..349M}, on the basis of a small parallax microlensing effect. The derived lens mass depends on the   lens and source distances. For example, $M_L\sim$13 $\msun$ for a source at the Galactic center distance of 8 kpc and a midway lens. 
The high likelihood of a BH nature was supported by a Bayesan statistical analysis that considered the mass functions of the disk and bulge stars \citep{2002ApJ...576L.131A}.
\citet{2005ApJ...633..914P} made a systematic analysis of 22 parallax microlensing events, taking into account various subtle effects that can lead to degenerate solutions and using a likelihood analysis to derive the masses.  MACHO-99-BLG-22 was found to be the strongest  BH candidate (78\%   probability) in their sample. In the following, we adopt their most likely values of $M_L\sim7.5$ $\msun$ and $D_L\sim4.8$ kpc.

\subsection{ MACHO-96-BLG-5 }

MACHO-96-BLG-5 is one of the two massive ($\sim$6~$\msun$) and dark (L$<1~\lsun$) lenses found in an analysis of the longest microlensing events in the MACHO survey \citep{2002ApJ...579..639B}. The lens mass-distance relation, assuming a source at the bulge distance of 8 kpc, gives most likely values in the ranges $M_L\sim3-16~\msun$ and $D_L\sim$0.5--2.2 kpc. The subsequent analysis of \citet{2005ApJ...633..914P} found a somewhat reduced probability of only 37\% for a BH nature.  A recent analysis of deep optical and near infrared obervations has strongly reduced the possible parameter space for a non BH lens in MACHO-96-BLG-5  \citep{2021ApJ...912..146A}.

A 10 ks long observation of MACHO-96-BLG-5 was carried out with the \cha\ ACIS-S instrument, but no X-rays were detected \citep{2005ApJ...631L..65M}, with a 99\% confidence level upper limits on the 0.3-8 keV absorbed flux of  $\sim5\times10^{-15}$ erg cm$^{-2}$ s$^{-1}$. Also a much longer (100 ks) observation with \xmm\ could not detect any X-ray emission  \citep{2006ApJ...651.1092N} and provided slightly worse upper limits.

\subsection{ MACHO-98-BLG-6 }

MACHO-98-BLG-6 is the second BH candidate proposed by \citet{2002ApJ...579..639B}.  With the assumption that the magnified star is at 8 kpc,  \citet{2007A&A...466..157S} derived $M_L=6^{+9}_{-4}$ $\msun$ and a distance $D_L$=1.9 kpc.  However,  according to the analysis of \citet{2005ApJ...633..914P} MACHO-98-BLG-6 has a very small probability of being a BH (2\%).

\subsection{ OGLE-2011-BLG-0310 }

OGLE-2011-BLG-0310 (also known as MOA-2011-BLG-332) is one of the five microlensing events for which an astrometric study has been carried out by \citet{2022arXiv220201903L}. However, these authors found an astrometric signal consistent with zero.    According to their analysis, OGLE-2011-BLG-0310  is most likely a white dwarf (65\% probability) or a NS (22\%) with mass of 0.78$^{+2.98}_{-0.68}$ $\msun$.

\subsection{OGLE3-ULENS-PAR-01, PAR-02, and PAR-05}

\citet{2016MNRAS.458.3012W} made an extensive search for  BH, NS and white dwarf candidates in the OGLE-III   database. They found  13 out of 59 parallax microlensing events consistent with a compact object lens. Their masses and distances were estimated assuming standard proper motions for stars in the Galaxy. Although it cannot be  excluded that some of them  are main sequence stars in the disk that move in parallel with bulge sources giving very small relative proper motions, a few interesting candidates were selected.
OGLE3-ULENS-PAR-02 and OGLE3-ULENS-PAR-05 are those with the highest estimated masses, consistent with being BH. Both of them have two possible solutions yielding slightly different parameters (see Table~\ref{tab-sources}).

\subsection{MOA-2009-BLG-260 }

MOA-2009-BLG-260 is an astrometric parallax event with mass $M_L=1.37^{+0.74}_{-0.60}$ \citep{2022arXiv220201903L}, most likely consisting of a NS (44\% probability). No \cha\ or \xmm\ observations of its position are available.

\subsection{ OGLE-00-BUL-43 }

\citet{2007A&A...466..157S} performed spectroscopy of 16 microlensing events to add radial velocity information and determine spectral types and estimated lens masses based on a  model of the Galaxy. 
OGLE-00-BUL-43 was found as a possible  BH  candidate, but no \cha\ or \xmm\ observations are available.

\begin{table*}[h!]
\caption{Isolated BH and NS candidates found with microlensing}
 \begin{tabular}{lccll}
   \hline
   \hline
      Source  &    Mass              & Distance         &     Comments &  References   \\
          &     ($\msun$)              &  (kpc)            &            &  \\
\hline 

\ibh\      &         7.1$\pm$1.3           &   1.58$\pm$0.18         &  $V_T$=45 km s$^{-1}$  & [S22]\\ 
               &       3.79$^{+0.62}_{-0.57}$    & [1.47-1.92]         & $V_T$=21-27 km s$^{-1}$,    100\% BH    & [L22]    \\ 
              &      2.12$^{+0.67}_{-0.54}$   &  [0.70-1.30]       & $V_T$=2-12 km s$^{-1}$,    44\%, BH,   50\% NS,   6\% WD &  [L22] \\ 
\hline 
MACHO-99-BLG-22    &   7.5       & 4.8 &   78\% BH,   7\% NS,   4\% WD  &   [M02,P05]    \\
\hline 
MACHO-96-BLG-5   &  6$^{+10}_{-3}$ & 1.3$^{+0.9}_{-0.8}$   & 37\% BH,  14\% NS,  19\% WD    &   [B02,P05]    \\
\hline 
 MACHO-98-BLG-6   &  6$^{+9}_{-4}$        & 1.9  &  2\% BH,   13\% NS,   26\% WD  &  [B02,P05,S07]    \\
\hline 
OGLE-2011-BLG-0310 &    0.78$^{+0.71}_{-0.39}$    &   4.3$^{+1.9}_{-1.6}$    &   5\% BH,  22\% NS, 65\% WD   &    [L22]\\ 
\hline 
OGLE3-ULENS-PAR-01 & 1.0$^{+1.8}_{-0.7}$    & 1.3$^{+1.1}_{-0.7}$   &       &  [W16]   \\
\hline 
OGLE3-ULENS-PAR-02 & 8.7$^{+8.1}_{-4.7}$ & 1.8$^{+1.1}_{-0.8}$ &  &   [W16] \\
                   & 9.3$^{+8.7}_{-4.3}$ & 2.4$^{+1.1}_{-1.0}$ &  &     \\
\hline 
OGLE3-ULENS-PAR-05 & 3.3$^{+2.7}_{-1.5}$  & 2.9$^{+1.1}_{-0.9}$    &   &    [W16]  \\
                   & 4.8$^{+4.0}_{-2.5}$  & 1.8$^{+1.2}_{-0.7}$   &    &      \\
\hline 
MOA-2009-BLG-260 & 1.37$^{+2.72}_{-1.16}$ & 5.0$^{+1.7}_{-1.3}$ &   14\% BH,   44\% NS,   38\% WD &  [L22]   \\
\hline 
OGLE-00-BUL-43 & 2.9$^{+6.1}_{-1.9}$ & 0.5 &    for disk lensed source  &    [S07]     \\ 
               & 2$^{+6}_{-1.5}$        &    0.8 & for  bulge lensed source     &      \\
\hline 
\end{tabular}
\label{tab-sources}

\textbf{References: }  [S22] \citet{2022arXiv220113296S}, [L22] \citet{2022arXiv220201903L}, [W16] \citet{2016MNRAS.458.3012W}, [P05] \citet{2005ApJ...633..914P}, [B02] \citet{2002ApJ...579..639B}, [S07] \citet{2007A&A...466..157S} , [M02] \citet{2002MNRAS.329..349M}.
 \end{table*}

\section{X-ray data analysis and results}
 \label{sec-results}

\cha\ observations are available for six of the microlensing events listed in  Table~\ref{tab-sources}:   \ibh,   MACHO-99-BLG-22, MACHO-98-BLG-5, OGLE-2011-BLG-0310, OGLE3-ULENS-PAR-05, and OGLE3-ULENS-PAR-01. Four of them have also \xmm\ observations, but these data have a worse sensitivity compared to the \cha\ ones and will not be considered in the following.  We report the results obtained with \xmm\ only for the two sources not observed with \cha : MACHO-98-BLG-6 and  OGLE3-ULENS-PAR-02. 
A log of the observations used in our analysis is given in Table~\ref{tab-obs}.

\subsection{\cha\  }

\cha\   data     reprocessing and analysis were performed with standard procedures using the most recent version of the \cha\ Interactive Analysis of Observation  (CIAO 4.14) and CALDB (4.9.6). 
Images and exposure maps were produced using {\sc fluximage} in the   energy range 0.5-7.0 keV. 
The script  {\sc srcflux} was used to estimate the radii of
the circular regions enclosing 90\% of the point spread function  at 1.0 keV.
Such radii range from 0.9 arcsec  on-axis  to 9 arcsec for the most    off-axis sources. 
The background for each observation was estimated using annular  regions centered on the source positions, with inner and outer radii of one and five times source extraction radius.
 These  source-free  background regions contain zero or, at most, a few counts.
The effective exposures used for the flux computation take into account the off-axis positions of the sources and have been computed taking the average exposure time inside the source extraction circle.

No X-ray sources were detected at the sky positions of the microlensing events.
Given the small number of detected counts, the upper limits on the  count rates (95\% c.l.) were  computed following \citet{1991ApJ...374..344K}. These limits were then converted to   fluxes  using the appropriate calibration files for each source and observation.  Such conversions depend on the assumed spectral shape and absorption. We used a power law spectrum with  photon index of 2 and  interstellar medium abundances from \citet{2000ApJ...542..914W}. The total hydrogen column densities in the directions of our targets are in the range  $[2-6]\times10^{21}$ cm$^{-2}$ \citep{2016A&A...594A.116H}. Given that the lenses are at intermediate distances between us and the Galactic bulge, these values can be regarded as upper limits. 
We inspected the run of extinction versus distance in the direction of each source derived by  \citet{2019ApJ...887...93G} and, based on the lens distance, we rescaled the total hydrogen  column  densities to obtain the $N_H$ values listed in Table  \ref{tab:chandra_ul}. These values were used to compute the unabsorbed fluxes. 
When HRC-I data were analysed, the upper limit on the net count rate was estimated in the energy range 0.1-10 keV, but the fluxes have been converted to the 0.5-7 keV band. 
All the results are reported in Table\,\ref{tab:chandra_ul}.
 
\subsection{\xmm\  }

An \xmm\ observation was targeted at the event OGLE3-ULENS-PAR-02 in September 2016.
We used the data obtained with the EPIC pn camera that was operated in Prime Full Window mode with the thin optical filter.
\xmm\ data were analysed with the Science Analysis Software (SAS)
adopting standard procedures. 
A faint source (4XMM\,J175723.4-284627 in the \xmm\ Serendipitous Source Catalog,  \citealt{2020A&A...641A.136W}) is detected at an angular distance of 6.6 arcsec from the position of OGLE3-ULENS-PAR-02. Its net count rate, measured using an extraction radius of 20 arcsec, is $(2.55\pm{0.75})\times10^{-3}$ counts s$^{-1}$ in the 0.2-12 keV range. The  error on its coordinates (0.98 arcsec statistical,  0.45 arcsec systematic, $1\sigma$ c.l.) is sufficiently small to exclude that this source is related to OGLE3-ULENS-PAR-02.

The instrument sensitivity  at the position of the gravitational lens is reduced due to  the presence of this nearby contaminating source.  Therefore, we conservatively assume as an upper limit the count rate of 4XMM\,J175723.4-284627. For a power law spectrum with photon index of 2 and $N_H=10^{21}$ cm$^{-2}$ this count rate corresponds to an unabsorbed 0.2-12 keV flux of $1.1\times10^{-14}$ erg cm$^{-2}$ s$^{-1}$ (or $7.1\times10^{-15}$ erg cm$^{-2}$ s$^{-1}$ in the 0.5-7 keV  range used above for \cha ).

The same \xmm\ observation covers also the sky position of the event MACHO-98-BLG-6, lying about 4.3 arcmin offset. For this event we analysed the EPIC MOS data (effective exposure T$_{exp}$=30.7 ks, MOS1), because its sky position lies very close to the gap between CCDs in the pn camera, hampering a proper determination of the upper limit.
We have measured a 95\%  upper limit to the count rate of $9.0\times10^{-4}$ counts s$^{-1}$ (0.2-12 keV), using the SAS script {\sc eupper}.
With the same assumptions on the spectrum used above, this translates into an upper limit to the unabsorbed flux of $1.78\times10^{-14}$ erg cm$^{-2}$ s$^{-1}$ ($1.14\times10^{-14}$ erg cm$^{-2}$ s$^{-1}$, 0.5-7 keV).

\begin{table*}[h!]
 \centering 
 \caption{X-ray observations of microlensing events. VF and F in column 5 indicate very faint and faint mode, respectively. }
 \begin{tabular}{llllllll}
   \hline
   \hline
      Source        & Satellite &  ObsID &  Date &        Instrument             &      Duration   & Effective  &       Off-axis    \\
                    &           &        &       &                                &                &    exposure   &     angle         \\     
                    &           &        &       &                                &     (ks)       &    (ks)   &       ($'$)          \\     
\hline
\ibh\                 &   Chandra & 8764        &    2008-05-13   &    ACIS-I (VF)   &    2.15 & 1.92      &    8.8    \\ 
                      &   Chandra & 13540       &    2011-11-01   &    ACIS-I (VF)   &    1.95 & 1.83      &    5.1    \\ 
                      &   Chandra & 21628       &    2019-07-29   &    ACIS-I (VF)   &    1.89 & 1.74      &   6.6     \\ 
                       &   Chandra   &    23951    &    2022-06-02  &    ACIS-I (VF)  &    5.29   &    4.84    &    5.1   \\ 
\hline
MACHO-99-BLG-22      &  Chandra &     3818     &   2003-08-01 & ACIS-S (VF)           &  9.7  & 9.36    &       0.0 \\  
\hline
MACHO-96-BLG-5      & Chandra    &   3789     &    2003-02-18  &      ACIS-S (F)    &    9.83 & 9.83   &       0.0 \\  
                    & Chandra   &  14677       &   2013-10-03   &      ACIS-I (F)    &    1.0     & 0.95 &    7.6       \\  
\hline
MACHO-98-BLG-6       & XMM &       0782770201    &    2016-09-03     &      EPIC MOS1        &    40      &   30.7   &      4.3       \\
\hline
OGLE-2011-BLG-0310    &  Chandra &  13542       &    2011-11-01            &  ACIS-I (VF)     & 1.95 &   1.864   &     8.7 \\  
                      &  Chandra &  13564       &    2011-11-01            &  ACIS-I (VF)     & 1.95 &   1.785  &    10      \\  
                      &  Chandra &  13565       &    2011-11-01            &  ACIS-I (VF)     & 1.95 &  1.744  &     5.9  \\  
                      &  Chandra &   23957      &        2021-03-18        &  ACIS-I (VF)     & 5.1 &   4.52   &     7.5    \\   
\hline
OGLE3-ULENS-PAR-01  & Chandra   &    7541      &      2007-10-23      &       HRC-I      &             1.19     &  1.17  &    13.8      \\
\hline
OGLE3-ULENS-PAR-02  & XMM &        0782770201   &       2016-09-03              &        EPIC pn       &    35          &    32   &      0.0        \\
\hline
OGLE3-ULENS-PAR-05  & Chandra    &   19723     &      2018-05-10      &       ACIS-I (VF)      &      19.8    & 19.8  &   0.0    \\
\hline
\end{tabular}
\label{tab-obs}
\end{table*}

\begin{table*}[h!]
 \centering 
 \caption{X-ray upper limits (95\% c.l.). 
 All count rates are in the energy range 0.5-7 keV, except \cha\ HRC-I rates, which are in the energy range 0.1-10 keV,  and \xmm\ count rates (0.2-12 keV). }
 \begin{tabular}{lllllll}
   \hline
   \hline
      Source            &   ObsID &     Count rate     &  Observed flux &  N$_H$      & Unabsorbed flux         &  Unabsorbed flux  \\ 
                          &                  &                        &  0.5-7 keV &                   &   0.5-7.0 keV        & 0.1-100 keV \\
                      &         &       (count s$^{-1}$)   &    (erg~cm$^{-2}$~s$^{-1}$) & (cm$^{-2}$)  & (erg~cm$^{-2}$~s$^{-1}$)      &  (erg~cm$^{-2}$~s$^{-1}$)  \\     
\hline
\ibh\                 &    8764            &    $<$0.0016    &  $<$2.20$\times$10$^{-14}$  &  10$^{21}$ & $<$2.57$\times$10$^{-14}$   & $<$6.76$\times$10$^{-14}$  \\ 
                      &    13540           &    $<$0.0016    & $<$1.93$\times$10$^{-14}$  & &  $<$2.25$\times$10$^{-14}$   & $<$5.92$\times$10$^{-14}$ \\ 
                      &    21628           &    $<$0.0017    &$<$3.16$\times$10$^{-14}$  &  &  $<$3.69$\times$10$^{-14}$   & $<$9.71$\times$10$^{-14}$ \\
                        &  23951            &    $<$0.00126     & $<$2.72$\times$10$^{-14}$  &    &      $<$3.17$\times$10$^{-14}$  &      $<$8.31$\times$10$^{-14}$             \\  
                      & combined          &    $<$0.00051      & $<$8.6$\times$10$^{-15}$  &   &      $<$1.0$\times$10$^{-14}$   &       $<$2.6$\times$10$^{-14}$            \\              
\hline
MACHO-99-BLG-22      &      3818       &    $<$3.2$\times$10$^{-4}$  & $<$1.99$\times$10$^{-15}$  & $3\times10^{21}$   &    $<$2.83$\times$10$^{-15}$  &    $<$7.43$\times$10$^{-15}$  \\  
\hline
MACHO-96-BLG-5       &  3789           &   $<$3.1$\times$10$^{-4}$   & $<$1.86$\times$10$^{-15}$  & 10$^{21}$ &   $<$2.17$\times$10$^{-15}$     &   $<$5.71$\times$10$^{-15}$        \\  
                                     &  14677            &     $<$0.0032          &  $<$5.63$\times$10$^{-14}$  & &  $<$6.57$\times$10$^{-14}$    &      $<$1.73$\times$10$^{-13}$     \\  
\hline
MACHO-98-BLG-6       &      0782770201      &   $<$9$\times$10$^{-4}$   & $<$9.77$\times$10$^{-15}$  & 10$^{21}$ &       $<$1.14$\times$10$^{-14}$  &     $<$3.0$\times$10$^{-14}$        \\
\hline
OGLE-2011-BLG-0310  & 13542  &  $<$0.0031   &  $<$4.37$\times$10$^{-14}$  & $5\times10^{21}$ &  $<$7.1$\times$10$^{-14}$   &  $<$1.9$\times$10$^{-13}$  \\  
                      &   13564      &    $<$0.0032          & $<$4.97$\times$10$^{-14}$  &  &  $<$8.0$\times$10$^{-14}$   &  $<$2.1$\times$10$^{-13}$ \\  
                      &   13565      &    $<$0.0026          & $<$3.60$\times$10$^{-14}$  & &  $<$5.8$\times$10$^{-14}$   &  $<$1.5$\times$10$^{-13}$  \\  
                      &    23957     &    $<$0.00066        & $<$1.42$\times$10$^{-14}$  & &  $<$2.3$\times$10$^{-14}$  &  $<$6.0$\times$10$^{-14}$     \\   
                     &  combined     &    $<$0.0009       & $<$1.54$\times$10$^{-14}$  & &  $<$2.5$\times$10$^{-14}$   & $<$6.6$\times$10$^{-14}$  \\  
\hline
OGLE3-ULENS-PAR-01   &    7541      &  $<$0.0178 & $<$4.37$\times$10$^{-13}$  &10$^{21}$   &  $<$5.1$\times$10$^{-13}$  &  $<$1.3$\times$10$^{-12}$  \\
\hline
OGLE3-ULENS-PAR-02  &    0782770201  &     $<$2.55$\times$10$^{-3}$  & $<$6.08$\times$10$^{-15}$  & 10$^{21}$   &   $<$7.1$\times$10$^{-15}$   &   $<$1.9$\times$10$^{-14}$        \\
\hline
OGLE3-ULENS-PAR-05  &   19723      & $<$0.00015  & $<$2.11$\times$10$^{-15}$  & $2\times10^{21}$ &   $<$2.76$\times$10$^{-15}$  &  $<$7.24$\times$10$^{-15}$  \\
\hline
\end{tabular}
\label{tab:chandra_ul}
\end{table*}

\begin{figure}[!h]
\begin{center}
\includegraphics[height=6.5cm,angle=0]{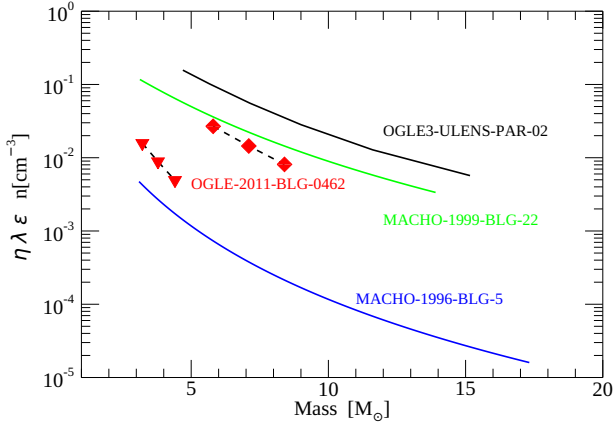} \\
\caption{\footnotesize{Constraints  on the  product of  BH efficiency  $\eta (\equiv L/\mdot c^2$)  and   ISM density derived from the X-ray flux upper limits of \ibh\ (red) and other BH candidates found with gravitational microlensing (see text for the definition of $\lambda$ and  $\epsilon$).  The red diamonds and triangles refer to the BH properties derived by \citet{2022arXiv220113296S} and \citet{2022arXiv220201903L}, respectively.  The curves for the other BH candidates are derived using for each mass value the corresponding distance and transverse velocity as given by the microlensing relations.}}
\label{fig-eta-M}
\end{center}
\end{figure}

\begin{figure}[!h]
\begin{center}
\includegraphics[height=6.0cm,angle=0]{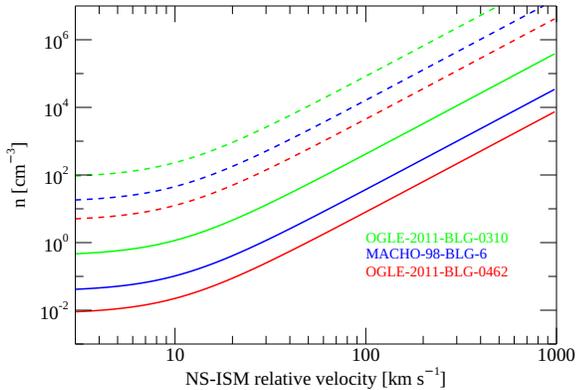} \\
\caption{\footnotesize{Constraints on the NS velocity and ISM density for candidate isolated NS found with gravitational microlensing.  A sound speed $c_s$=10 km s$^{-1}$ has been assumed}. The regions above the solid lines are excluded by the upper limits on the soft X-ray flux obtained  with \cha , while those above the dashed lines are excluded by the INTEGRAL limits in the  hard  X-ray range (17-60 keV). }
\label{fig-NS}
\end{center}
\end{figure}

\begin{figure}[!h]
\begin{center}
\includegraphics[height=6.5cm,angle=0]{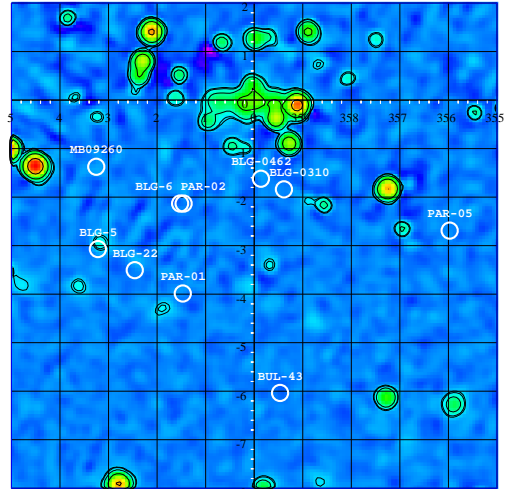} \\
\caption{\footnotesize{Image in the 17-60 keV range obtained with the INTEGRAL IBIS instrument (adapted from \citet{2022MNRAS.510.4796K}). The image is in Galactic coordinates and covers a region of  about 10$\times$10 deg$^2$. The positions of the BH/NS candidates found with gravitational microlensing are indicated by circles with radius of 10 arcmin.  The   source on the border of the MACHO-96-BLG-5 circle,  IGR J18044--2739, is identified with a magnetic cataclysmic variable \citep{2012A&A...544A.114M}. Its coordinates are incompatible with those of the microlens source.}}
\label{fig-integral}
\end{center}
\end{figure}

\section{Discussion}
\label{sec-disc}

The accretion-powered luminosity of an isolated compact star of mass $M$ moving with velocity $V$ in the interstellar medium of density $n$ depends on the mass accretion rate $\mdot$, that, in the Bondi-Hoyle description, is given by

\begin{equation}
\mdot = 4\pi n m_p    \frac{(GM)^2}{(V^2 + c_s^2)^{3/2}}  \lambda   ~~{\rm  g~s^{-1}}.
\label{eq-mdot}
\end{equation}
  
\noindent 
Here $G$ is the gravitational constant, $m_p$ the proton mass, and $c_s$  the sound speed,  which we will neglect in the following because, for the most common phases of the ISM, it is much smaller than the typical space velocity of isolated NS and BH. 
The parameter $\lambda$ accounts for the fact that not all the matter crossing the Bondi-Hoyle radius is  accreted. As discussed in \citet{2018MNRAS.477..791T}, its value is   subject to a large uncertainty. 
Therefore, we keep it as an explicit scaling factor in all the results given below  and in Fig.~\ref{fig-eta-M}. 

According to the analysis of \citet{2022arXiv220201903L}, there is a non negligible probability that \ibh\ is a NS rather than a BH; also other objects listed in Table~\ref{tab-sources} are most likely NS. Therefore, in the following we  discuss both the BH and NS case.

\subsection{Black hole}

\citet{2022arXiv220113296S}, derived for  \ibh\  a transverse   velocity of 45 km s$^{-1}$.  We can thus neglect $c_s$ in eq.~\ref{eq-mdot} and, with the 7.1 $\msun$ mass given by these authors, estimate an  accretion rate of  $2\times10^{11} ~n\lambda$ g s$^{-1}$. This is actually an upper limit because it does not take into account the
unknown  component of the velocity along the line of sight.  
Then, for the BH distance $d$=1.58 kpc,  the upper limit  on the X-ray flux derived above, $F_X<8.9\times10^{-15}$ erg cm$^{-2}$ s$^{-1}$,    implies a luminosity

\begin{equation}
L <  \frac{F_X}{\epsilon} 4\pi d^2   = 2.7\times10^{30} / \epsilon   ~~~~{\rm  erg~s^{-1}}, 
\label{eq-Lul}
\end{equation}

\noindent
and an efficiency of conversion of gravitational energy to electromagnetic radiation 

\begin{equation}
\eta\equiv L/(\mdot c^2) < \frac{0.015}{\epsilon\lambda}   
\left(\frac{n}{1~{\rm cm^{-3}}}\right)^{-1},
\label{eq-eta}
\end{equation}
  
\noindent
where we introduced the factor   $\epsilon$ to  account for the fraction of accretion luminosity falling outside the observed  X-ray  band. Assuming that the  luminosity is emitted in the 0.1-100 keV range with a power-law spectrum of photon index $\alpha$=2, the 0.5--7 keV range used above corresponds to $\epsilon$ = $F_{0.5-7}/F_{0.1-100}$ = 0.38    ($\epsilon$= 0.2 / 0.34 for $\alpha$ = 1.5 / 2.5).

\citet{2022arXiv220201903L} found for the BH case a most likely  mass of  3.8 $\msun$  and a  transverse velocity in the range   21-27 km s$^{-1}$. 
The mass accretion rate has a  strong dependence on the BH velocity.
Thus, despite the smaller   mass, the limit on $\eta$ obtained using these values  is  more constraining than that derived above.
This is shown in Fig.~\ref{fig-eta-M}, where the upper limit on the product $\eta \lambda \epsilon n$ is plotted as a function of the BH mass for both the \citet{2022arXiv220113296S} and \citet{2022arXiv220201903L}   parameters.

We plot in  Fig.~\ref{fig-eta-M} also the constraints obtained in a similar way from our   upper limits on the X-ray fluxes of the other candidates of Table~\ref{tab-sources} that are likely to be BH.
To derive the limits  for these sources, which, contrary to \ibh , do not have an astrometric parallax measurement,  we took into account the distance  dependence of the mass and transverse velocity.
Our constraints for  MACHO-96-BLG-5
agree  with those derived by \citet{2005ApJ...631L..65M} using the same data.
These constraints are below those of \ibh , but note that, besides having a large distance uncertainty, after the analysis of \citet{2005ApJ...633..914P} MACHO-96-BLG-5 is no more  considered a strong BH candidate.

\subsection{Neutron star}

The expected  accretion luminosity for a NS of mass $M_{NS}$ and radius $R_{NS}$  is  

\begin{equation}
L_{NS}  = \frac{G M_{NS}}{R_{NS}} \mdot =\frac{(GM_{NS})^3}{R_{NS}V^3} 4\pi m_p n \lambda,
\end{equation}
 
\noindent
which, for $M_{NS}=1.4~ \msun$ and   $R_{NS}$=10 km,   
gives a luminosity

\begin{equation}
L_{NS}  =  1.3\times10^{32} 
 \left(\frac{10~{\rm km~s^{-1}}}{V}\right)^3 
n \lambda~~{\rm erg~s^{-1}}.
\end{equation}

For the small  distance (0.7--1.3 kpc at 99\% c.l.) and transverse velocity (2--12 km s$^{-1}$)  derived for \ibh\ in the NS case \citep{2022arXiv220201903L}, the resulting flux  is  in the broad range
$F_X=[0.6-140]\times10^{-12}  n \lambda / d^2_{kpc}$ erg cm$^{-2}$ s$^{-1}$.
This expected flux is close to, or higher than, the  upper limit  of   $\sim10^{-14}$ erg cm$^{-2}$ s$^{-1}$ derived above, thus possibly disfavoring a NS nature. 

Taking booth $\lambda$ and $d_{kpc}\sim$1, implies that the NS is in a low density ( $n\lesssim10^{-2}$ cm$^{-3}$) environment, but unfortunately, no strong conclusions can be drawn because the radial component of the NS is unknown.
Isolated NS have  relatively high space velocities.   \citet{2005MNRAS.360..974H} found that radio pulsars have a velocity distribution consistent with a Maxwellian with rms $\sigma$=265 km s$^{-1}$, while a  more recent analysis  found that the sum of two Maxwellian distributions with  $\sigma_1$=128 km s$^{-1}$ and $\sigma_2$=298 km s$^{-1}$ gives a better description of the pulsar velocity distribution \citep{2020MNRAS.494.3663I}. 

We show  in  Fig.~\ref{fig-NS} how the upper limits on the X-ray flux we derived for the microlensing NS candidates translate into constraints on  the ISM density as a function of  their space velocities. For all the candidates we assumed $M_{NS}=1.4~ \msun$,  $R_{NS}$=10 km,  the best fit distances given in Table~\ref{tab-sources},  $\lambda=1$,  and an ISM sound speed of $c_s$= 10 km s$^{-1}$.  For example, if   \ibh\ has a space velocity below 100  km s$^{-1}$, it cannot be in a region of the ISM with density higher than  $\sim$10 cm$^{-3}$. 
This excludes that \ibh\ is either inside or within the envelope of a molecular cloud.

\subsection{Limits in the hard X-ray band}

If these compact objects are in dense molecular clouds, their flux can be severely absorbed in the softest part of the energy range explored with \cha\ and \xmm . It is therefore interesting to search also for X-ray emission in the hard X-ray range, that is unaffected by interstellar absorption. All the microlensing events considered here are in the direction of the Galactic bulge, a region that has been extensively observed with the INTEGRAL satellite.   The positions of the microlensing candidates are marked in Fig.~\ref{fig-integral} on a sky map    obtained with the INTEGRAL IBIS instrument \citep{2003A&A...411L.131U} in the 17-60 keV energy range.   This image has been derived by summing observations carried out from 2002 to 2017 and corresponds to a total net exposure time of more than 13 millions of  seconds \citep{2022MNRAS.510.4796K}. The resulting limiting sensitivity is $\sim2\times10^{-12}$ erg cm$^{-2}$ s$^{-1}$.

Since no hard X-ray sources are detected at the positions of  the microlensing events, we used this flux value  to derive the limits shown by the dashed lines in Fig.~\ref{fig-NS}. Although these limits are above those obtained in soft X-rays, they do not  depend on the absorption and can be applied also for sources inside dense molecular clouds.

\section{Conclusions}
\label{sec-conc}

We derived upper limits on the X-ray flux from   \ibh .
This is currently the strongest BH candidate selected through astrometric  gravitational microlensing \citep{2022arXiv220113296S}, although there is some tension in its parameters as derived by different groups and a NS nature cannot be completely excluded \citep{2022arXiv220201903L}.
 Our limits are consistent with       an accreting isolated BH with low radiative efficiency, as observed in quiescent X-ray binaries and in AGNs, and do not  support  a NS nature for the lens in \ibh . However, we cannot exclude the possibility of a NS in a relatively low density environment and/or with a space velocity much larger than the projected value obtained from the gravitational microlensing analysis. 

Using  \cha\, \xmm\ and INTEGRAL data, we derived upper limits on the  soft and hard X-ray fluxes also for several other candidate compact objects found with gravitational microlensing. The limits reported here are the best ones currently available for these objects, but the resulting   constraints on the  physical parameters governing the accretion process depend on several poorly known factors.
Among these, a major role is played by the space velocity of the compact object, since even a factor of two yields  a difference of about one order of magnitude in the expected accretion rate. In fact,  the relative positions of the limits plotted  in Fig.~\ref{fig-eta-M} are determined mainly  by the different velocities of these objects.  

Further observations will provide better estimates of the masses and distances of these candidate collapsed objects and  possibly  discriminate between   BH and  NS, but the radial velocity component of dark lenses cannot be directly measured. This problem will always affect the interpretation of X-ray observations of  individual sources.   
However,  many more isolated BH and NS will be discovered with microlensing surveys in the future. Thanks to the  availability of a large sample of reliable candidates observed with  high-sensitivity X-ray telescopes, it will be possible to perform statistical analysis exploiting our best knowledge of the distribution of spatial velocities, thus deriving meaningful constraints on the physics of accretion from the ISM onto isolated compact objects.

\acknowledgments 
We thank the anonymous referee for useful suggestions. We acknowledge financial support from the Italian Ministry for University and Research through grant 2017LJ39LM ``UnIAM'' and the  INAF ``Main-streams'' funding grant (DP n.43/18). 
GP acknowledges funding from the European Research Council (ERC) under the European Union’s Horizon 2020 research and innovation programme (grant agreement No 865637).
This research has made use of data obtained from the \cha\ Data Archive and software provided by the \cha\ X-ray Center (CXC) in the application package CIAO.
This research is based on observations obtained with \xmm, an ESA science mission with instruments and contributions directly funded by ESA Member States and NASA. The \xmm\ data were downloaded by means of the \xmm\ Science Archive (XSA v14.1) and of the High Energy Astrophysics Science Archive Research Center (HEASARC), a service of the Astrophysics Science Division at NASA/GSFC.

\bibliographystyle{aasjournal} 

\end{document}